\documentstyle[twoside,fleqn,jhuwkshp,psfig]{article}
\newcommand{\etal}{{\it et al.} }

\begin{document}

\title{MODELS FOR THE UV-X EMISSION IN AGN: INFLUENCE OF DIFFERENT PARAMETERS}
 
\author{S.Coup\'e
\address{DAEC, Observatoire de Paris-Meudon, 92195 Meudon Cedex, France},
M-C. Artru \address{CRAL, ENS Lyon, France},
S. Collin $^{a}$, 
B. Czerny \address{Copernicus Astronomical Center, Bartycka 18, 00-716 Warsaw,
Poland},
A-M. Dumont $^{a}$}

\begin{abstract}

We present the results of computations performed with a
new photoionisation-transfer code designed for hot Compton thick 
media.
Presently the code solves the transfer of the continuum with the 
Accelerated Lambda Iteration method (ALI) and that of the lines 
in a two stream Eddington approximation, without using the local 
escape probability formalism to approximate the line transfer.
We show the influence that the approximations made in the treatment 
of the transfer and of the atomic data can have on the broad band 
UV-X spectrum and on the detailed spectral features (X-ray lines 
and ionization edges) emitted/absorbed by an X-ray irradiated medium.
This transfer code is coupled with a Monte Carlo code which allows 
to take  into account direct and inverse Compton diffusions, and 
to compute the spectrum emitted up to hundreds of keV energies.
The influence of a few physical parameters is shown, and the 
importance of the density and pressure distribution (constant 
density, pressure equilibrium) is discussed. 
\end{abstract}

\maketitle

\section{The codes}

Using a photoionisation-transfer code specially designed for hot Compton 
thick media
($T \geq 10^4 K$, Thomson depth $\tau_{es}$ can be as large as a few hundreds),
we are able to compute the thermal and ionisation structure and the 
spectrum emitted and reflected by a slab of gas illuminated
on one side or on both sides by a given spectrum in a plane-parallel geometry
(for a description of the code, see Dumont \etal ~\cite{dum00} and Dumont \&
Collin ~\cite{dum01}). Contrary to the other photoionisation codes which
use the escape probability (EP) formalism, our code
solves the transfer of the continuum with the Accelerated Lambda Iteration 
(ALI) procedure and that of the lines with a two stream approximation (ALI is 
being implemented presently for the lines). It is coupled with a Monte-Carlo 
code which allows to take into account Compton and inverse Compton diffusions,
and to compute the spectrum emitted up to hundreds of keV energies, in any 
geometry.
\par
The gas composition includes 10 elements and 102 ions. H is treated as a 
6-level atom, H-like and Li-like as 5-level atoms, He-like ions as 8-levels 
atoms and matrix inversions are performed to compute the levels populations.
The other ions are treated roughly as two level atoms plus a continuum.
All collisional and radiative processes (including ionizations from and
recombinations onto excited levels) are taken into account. Recombinations onto
upper levels ($n > 6$ for H, $n > 5$ for H- and Li-like ions, $n > 
8$ for He-like ions) are added to our highest excited level.

\section{Results}

We consider here two cases :
\begin{itemize}
\item the density is constant across the slab
\item the pressure is constant across the slab
\end{itemize}
In the present examples the incident continuum is a power law with a slope
of unity (in flux, $F_\nu$) from 0.1 eV to 100 keV. We take cosmic abundances.
The density at the surface of the slab is $10^{12}$cm$^{-3}$ and the incident 
flux is equal to $8 \ 10^{13}$ erg cm$^{-3}$ s$^{-1}$ (corresponding to an 
ionisation parameter $\xi = L/(n_H R^2) = 1000$ erg.cm.s$^{-1}$) in constant 
pressure and constant density cases.

\subsection{Influence of the transfer treatment} 

In Dumont \etal ~\cite{dum00}, it was shown that the neglect of the returning 
radiation often used in photoionisation codes, i.e. the ``outward only'' 
approximation, leads to strong temperature errors, in particular at the 
illuminated face of the slab when the slab is thick ($\tau_{es} \geq 0.1$).
\par
\begin{figure}[tbp] % fig.1
\vspace{10pt}
\centerline{\psfig{file=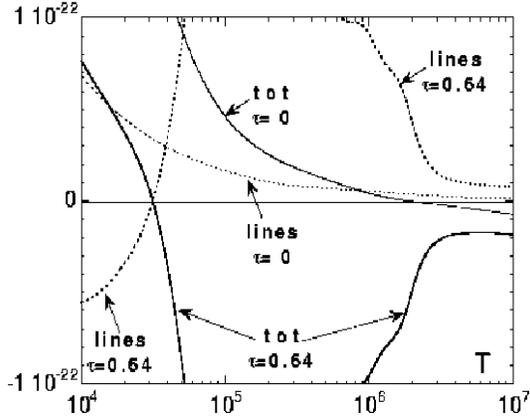,width=3.0in,height=2.5in,angle=270}}
\caption{Radiative rate cooling-heating curves (full lines) 
and net cooling curves due to lines (dashed lines) for different optical 
thicknesses in the constant pressure case ($\tau_{es}=0$ and equilibrium 
temperature $T_{eq}=2\ 10^6 K$, $\tau_{es}=0.64$ and $T_{eq}=3.5\ 10^4 K$). 
One can see that the line cooling which dominates the energy balance becomes 
negative close to the equilibrium temperature.}\label{fig:gppcst}
\end{figure}

 
One other difference between EP and transfer lies in the fact that the line
emissivity can be negative whereas it is impossible in EP (see 
~\cite{dum00}). This translates into a difference in the energy balance, 
which can have important consequences on the equilibrium temperature
(see Fig.~\ref{fig:gppcst} for an example in the constant pressure case).

\subsection{Influence of the atomic data}

\begin{figure}[tbp] % fig.2
\vspace{10pt}
\centerline{\psfig{file=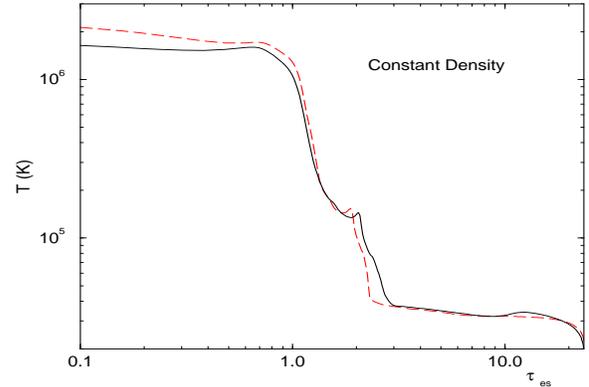,width=3.0in,height=2.0in}}
\caption{Temperature profile versus optical thickness, for
a treatment with only multi-level H-like ions (dashed lines), and
with H-like, He-like and Li-like multi-level ions (full lines)
for a constant density model.}\label{fig:profTncst}
\end{figure}

\begin{figure}[tbp] % fig.3
\vspace{10pt}
\centerline{\psfig{file=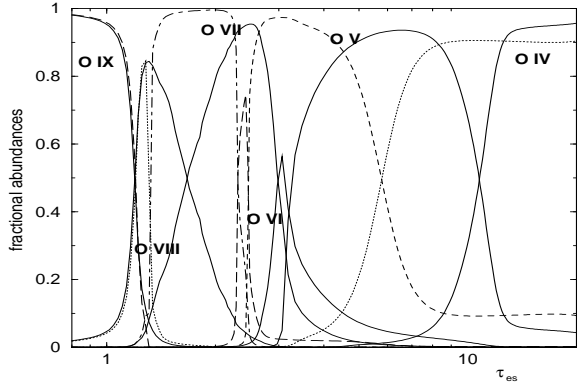,width=3.0in,height=2.0in}}
\caption{Fractional abundances of oxygen versus optical thickness for 
the same two treatments of the atomic data.
As in Fig.~\ref{fig:profTncst}, the example shown here is for a constant 
density model.}\label{fig:dioOncst}
\end{figure}

The model atoms are important not only for the thermal balance, but also
for the ionization equilibrium. Fig.~\ref{fig:profTncst} shows 
the temperature profile and Fig.~\ref{fig:dioOncst} the
fractional abundances of oxygen versus optical thickness, for two 
different treatments of the atomic data, in the constant density case. 
When only H-like ions are treated with 
multi-level atoms, the temperature is higher than in the treatment with 
H-like, He-like and Li-like multi-level ions for an optical thickness
smaller than about 2 and then is lower, for larger optical thicknesses.
\par
Treating H-, He-, Li-like ions as multi-level atoms produce more lines 
than with only H-like multi-level atoms; all these added lines cool the medium
which leads to a cooler temperature in the case of H-, He-, Li-like multi-level
ions. The temperature profile declines faster in the H-like description than
in the 'total' description between $\tau_{es} =2$ and $\tau_{es}=3$ because 
in that
place, photoionization from the excited levels of Li-like ions are as important
as those from the ground level in contradistinction to H-like and He-like
ions where the photoionisation from excited levels is always negligible.
When looking at the fractional abundances of oxygen, one sees that O VIII 
is present deeper in the cloud in the description with H-, He-, Li-like multi
level atoms than in the treatment with only H-like multi level atoms. 
This is due to an approximate treatment of losses due to recombination
radiation in the two-level atoms.      

\begin{figure}[tbp] % fig.4
\vspace{10pt}
\centerline{\psfig{file=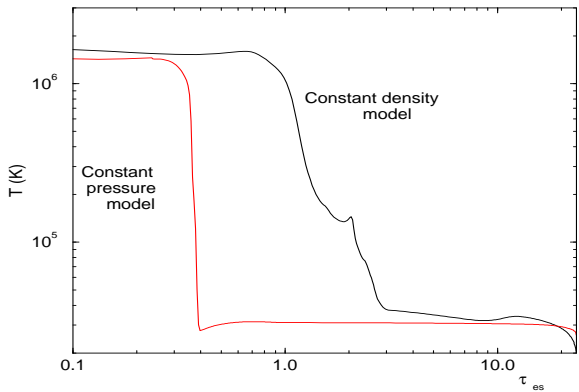,width=3.0in,height=2.0in}}
\caption{Temperature profile versus optical thickness for
the constant density and the constant pressure cases. The temperature 
decreases abruptly in the constant pressure case, when the soft X-ray photons
are absorbed.}\label{fig:profTpcst}
\end{figure} 

\begin{figure}[tbp] % fig.5
\vspace{10pt}
\centerline{\psfig{file=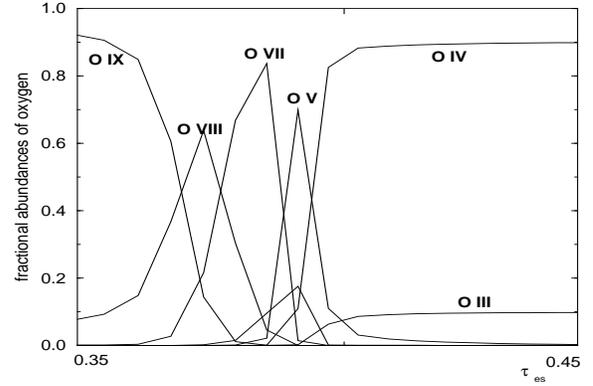,width=3.0in,height=2.0in}}
\caption{Fractional abundances of oxygen versus optical thickness for the 
constant pressure case. Note that, in the constant density case 
(fig~\ref{fig:dioOncst}), only highly 
ionized species are present in the regions contributing to the emission, while
in the constant pressure case, both high and low ionization species are
present.}\label{fig:dioOpcst}
\end{figure} 

\begin{figure}[tbp] % fig.6
\vspace{10pt}
\centerline{\psfig{file=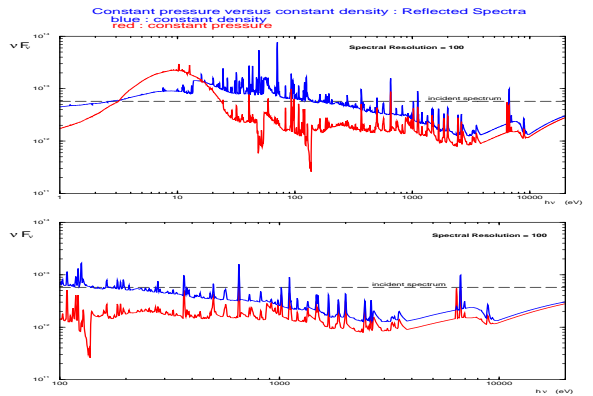,width=3.0in,height=2.0in}}
\caption{Comparison of the reflected spectrum between 1 eV and 25 keV
in a constant density and a constant pressure cases.}\label{fig:speI}
\end{figure} 

\subsection{Comparison of constant density and constant pressure models}

\begin{figure}[tbp] % fig.7
\vspace{10pt}
\centerline{\psfig{file=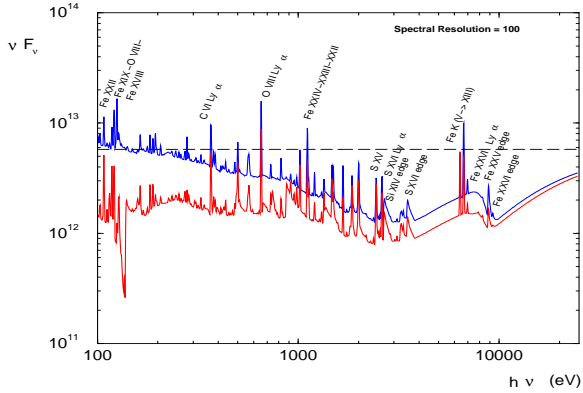,width=3.0in,height=2.0in}}
\caption{Zoom of Fig.~\ref{fig:speI} between 100 eV and 20 keV. A few lines 
and edges are labelled in the figure, but others are identified in Table 
~\ref{tab:ew}.}
\label{fig:speII}
\end{figure}

The slab being illuminated with the same irradiating flux, we compare the 
results in the constant density and in the constant pressure cases. 
The temperature profiles and fractional abundances of oxygen 
are shown on Figs.~\ref{fig:profTncst},~\ref{fig:profTpcst} and 
~\ref{fig:dioOncst},~\ref{fig:dioOpcst} respectively. 
At first sight, one sees that the thermal and ionization structure are 
completely different in the two different cases, owing to the phase change of 
the photoionized gas in the constant pressure case (see Krolik
\etal~\cite{kro81}) when soft X-ray photons have been absorbed. The 
temperature decreases abruptly and consequently the density increases in the 
constant presssure case. Accordingly, the emission-reflection spectra are
different. Fig.~\ref{fig:speI} shows the
reflected spectra in the constant density and the constant pressure cases, and 
Fig.~\ref{fig:speII} is a zoom of Fig.~\ref{fig:speI} between 0.1 and 20 keV
on which a few lines and edges are identified. 
Some line equivalent widths (EWs) of interest, in the constant 
density and constant pressure cases, are given in Table 
~\ref{tab:ew}. We recall that the EWs are those measured only on the 
reflected spectra with no dilution due to the incident spectrum.
\par
Due to the absence of layers with an 
intermediate temperature in the constant pressure case, the ``intermediate''
ions are almost absent. As a consequence, the iron lines correspond to
both high and low ionization species, in contradistinction to the constant 
density case. Finally, due to the absence of intermediate temperature layers 
giving rise to soft X-ray reflection, the reflected spectrum in the constant 
pressure case displays a narrower ``Blue Bump'' than in the constant density 
case. 

\section{Foreseen improvements}

Presently we are implementing the ALI method in the transfer of the lines which
are the most difficult to converge.
\par
We intend to treat all ions as multi-level atoms and to add levels to those
already treated like this. We will take into account some forbidden lines
which could be important for the cooling of the medium in some places.
\par
Though there are still many improvements to perform, we already use theses 
codes to describe AGN spectra in the UV and X-ray range, in their validity 
range. In particular, the code is used to compute the spectrum emitted by
an irradiated slab in hydrostatic equilibrium (see R\'o\.za\'nska \etal in
this workshop).

\begin{table}[hbt]         % table 1
\caption{\centerline{Measured Equivalent Widths (reflected spectra)}}
\label{tab:ew}
{\scriptsize{
\begin{center}
\begin{tabular}{@{}llll@{\extracolsep{\fill}}r}
\hline
Ion & $h\nu$(eV) & EW$^a$(eV) &  EW$^b$(eV) \\
\hline
\\
Fe XXVI $Ly\alpha$ & 6957.7 & 28.06 & 17.15 \\
Fe XXV & 6667.7 & 237.7 & 145.6 \\
Fe K $(V \rightarrow XIV)$ & 6400.0 & 6.362 & 180.6 \\
S XVI  & 3342.8 & 4.594 & 6.056 \\
S XVI $Ly\alpha$ & 2611.3 & 42.90 & 47.16 \\
S XV  & 2450.3 & 32.28 & 42.51 \\
Si XIV $Ly\alpha$ &  1999.8 & 27.18 & 30.67 \\
Si XII & 1865.0 & 4.892 & 7.637 \\
Fe XXIV & 1673.2 & 27.44 & 21.00 \\
Fe XXIV & 1495.6 & 16.01 & 10.57 \\
Mg XI & 1343.3 & 6.034 & 11.19 \\
Fe XXII-XXIII & 1127.1 & 3.698 & 6.986\\ 
Fe XXIV &1110.0 & 35.03 & 26.44 \\
Ne X $Ly\alpha$ &1020.3 & 13.46 & 16.30 \\
Fe XVII & 821.10 & 4.957 & 4.025 \\
Fe VII &  729.33 & 1.951 & 2.978 \\
O VIII $Ly\alpha$ & 653.35 & 26.37 & 32.75 \\
Fe XXIV & 563.57 & 2.466 & 2.001 \\
N VII $Ly\alpha$ &  500.14 & 3.895 & 5.429 \\
C VI $Ly\alpha$ &  367.36 & 5.263 & 9.009 \\
Mg X & 194.61 & 1.079 & 0.468 \\
O VIII & 163.33 & 0.744 & 1.360 \\
Fe XVIII & 127.82 & 0.285 & 1.578 \\
O VIII &  120.98 & 1.577 & 2.706 \\
Fe XIX &  118.08 & 0.711 & 2.327 \\
Fe XXII &  107.81 & 0.878 & 3.376 \\
\hline
$^a$ constant density case \\ 
$^b$ constant pressure case. \\ 
\end{tabular}
\end{center}}}
\end{table}

%do not change this

%do not change this
\normalsize

\end{document}